# Building an OSS Quality Estimation Model with CATREG

Jie Xu, Luiz Fernando Capretz
Department of Electrical and Computer Engineering
University of Western Ontario
London, N6A 5B9 Canada

Danny Ho
NFA Estimation Inc.
Richmond Hill, L4C 0A2 Canada

Correspondence should be addressed to Jie Xu, jxu89@uwo.ca

*Abstract*—**Open Source Software (OSS) has been a popular form in software development. In this paper, we use statistical approaches to derive OSS quality estimation models. Our objective is to build estimation models for the number of defects with metrics at project levels. First CATREG (Categorical regression with optimal scaling) is used to obtain quantifications of the qualitative variables. Then the independent variables are validated using the stepwise linear regression. The process is repeated to acquire optimal quantifications and final regression formula. This modeling process is performed based on data from the OSS communities and is proved to be practically valuable.**

## I. INTRODUCTION

For open source software, there are very few researches on quality estimation model, although some analyses have been carried out for the quality issues in this area. Aberdour maintained that sustainable community, code modularity, project management, and test process management were key areas in OSS quality management [1]. Koch and Neumann conducted a survey among hundreds of OSS projects, and analyzed the relationships among process metrics, product metrics and faulty classes [2]. The results were obtained both at class level and project level, but they only included qualitative comparisons. Another study adopted the defect content estimation approach from closed environment, i.e. using OO design metrics to derive the number of defects in modules [3].

Theoretically, all methods to derive software quality estimation models in industry or closed environment can be replicated for OSS projects. However, OSS projects have many unique characteristics in development compared with software projects in industry. Extra efforts are required to determine effective software metrics for quality estimation. Then suitable approaches can be adopted to build estimation model for quality management purpose.

We decide to concentrate on using quality predictors at project level to estimate the number of defects in the project since they can be obtained at early stages of software development. In current practices of software project management, the number of defects is still a key issue to trace, fix and manage.

The remainder of this paper is organized as follows. In Section II, modeling methodology is discussed to establish estimation models. Section III describes the details of data preparation. Empirical results are presented to verify the process in Section IV. Finally there are the conclusions and future work in Section V.

## II. MODELING METHODOLOGY

Statistical techniques are still among the most popular ones for modeling purposes in software engineering. Applicable regression techniques need to be explored to fit the data and establish a model. First the distribution characteristics should be analyzed to determine the form of the formula. Then corresponding regression techniques need to be applied to calculate the parameters in the function.

Some problems have to be solved for building the regression model. The form of the regression model should be decided first. The number of defects found in the software lifecycle is regarded as the dependent variable for this estimation task, which may have various forms of relationships with the selected predictors. In this paper, it can be treated as a linear one with appropriate transformation to the variables [4].

Then certain regression technique should be chosen to derive the parameters in the function. Although the function can take a linear form, ordinary linear regression is not applicable here because many predicators (independent variables) are categorical. Usually dummy binary variables have to be designed to apply traditional linear regression, but the results would be hard for interpretation and impossible for further recalibration. A special approach named CATREG (Categorical regression with optimal scaling using alternating least squares) is suitable to assign numerical values to those categorical variables and obtain the final regression formula [5]. The rationale behind it is transforming the categorical variables according to the optimal scaling levels (nominal or ordinal) and optimizing the quantifications following the least square criterion [6]. Using CATREG, the quantifications are achieved at the same time the regression is done.

With the results from CATREG, we still tend to verify the statistical significance of the predictors. For numeric variables, a linear transformation is made during CATREG. Consequently, CATREG is equivalent to a standard linear regression when the qualitative predictors are substituted by the transformed values (optimal scaling). As a result, traditional regression techniques like stepwise linear regression can be applied by assigning the obtained optimal scaling values to the qualitative independent variables. Therefore stepwise linear regression is performed to decide what independent variables are valid to enter into the regression model. The process can be repeated until satisfactory results are obtained.

### III. DATA PREPARATION

#### A. Data Source

SourceForge.net is the largest OSS development website in the world. As of February 2009, there have been more than 230,000 OSS projects registered to use the development services and more than 2 million registered users involved in the development activities. Many researchers who are interested in exploring the inherent characteristics of OSS projects have chosen it as the primary data source.

SourceForge data has been shared with the University of Notre Dame for research purposes and it consists of more than 100 tables in the data dumps. A project named FLOSSmole (Collaborative collection and analysis of free/libre/open source project data) has been developed to share data about OSS projects to the public domain [7]. Web crawling of the most popular OSS hosts, including SourceForge, has been performed on monthly basis to collect data from those websites. We concentrated on OSS projects hosted on SourceForge and extracted related project information of status, ranks, and developers from the above sources.

#### B. Data Collection and Integration

Based on popularity, status and other criteria, we selected 1571 OSS projects from SourceForge. Some information of the projects was accumulated from FLOSSmole and the data dumps from Notre Dame, but it did not cover all project characteristics that were related to software quality. Thus we designed a questionnaire to collect related information (Appendix A), which comprised of 22 multiple-choice questions. Some questions about product complexity were adopted from those of COQUALMO [8].

The survey was conducted by sending the questionnaire to the project administrators by email. The number of responses was not so satisfactory. Only 278 valid responses were received out of 1571 sent emails. Most of the sent emails might be filtered as junk emails or were neglected. Only a few of the respondents clearly showed no intent to do it.

We decided to use function points to measure size. Since we did not have the entries to calculate function points directly, we counted logical lines of code according to language types first for each project, and then applied backfiring method to obtain function points [9].

Finally 194 OSS projects were kept for doing experiments after removing outliers and data points with missing information. The data items contained the responses of 22 questions, size of the software, duration of the project, team size, the number of defects, etc.

### IV. EMPIRICAL RESULTS

To apply the modeling methodology discussed before, first we transformed those quantitative variables by natural logarithm to make them conform to normal distributions (Appendix B).

CATREG was performed to the 22 qualitative (22 questions) and 3 quantitative independent variables, with the number of defects as the dependent variable.

The CATREG summary was displayed in Table 1, with adjusted R-square being 0.471 and p-value 0.

We did not list the coefficient results of CATREG since there were so many independent variables. The results showed that many of the derived coefficients were statistically significant (p-value > 0.05), which meant that some of the independent variables should be excluded from the regression model.

Then we applied stepwise linear regression to the data, with the resulted quantifications for the respective independent variables. The summary of the first-round stepwise was presented in Table 2. Only the last step (No. 9) of the stepwise regression was listed and the results of the previous stepwise steps were omitted here for better demonstration. The results showed the final model included only 9 predictors, with adjusted R-square being 0.523 and p-value 0.

We also examined the coefficients of the final model and found all of them were significant (p-value < 0.05, Table 3).

The independent variables entered into the regression model were Q2, Q3, Q9, Q10, Q11, Q17, Q18, Ln(FP) and Ln(Duration). Because CATREG would result in different quantifications if the number of variables is different in the regression process, we had to perform CATREG for another round but only including the above independent variables and the dependent variable.

The model summary of the second-round CATREG was displayed in Table 4, with adjusted R-square being 0.501 and p-value 0. The R-square was better than that of the first-round, even though with less predictors in the model.

TABLE 1. Model summary of first-round CATREG

|  | Multiple R | R Square | Adjusted R Square |
|---|---|---|---|
| Standardized Data | 0.766 | 0.586 | 0.471 |
| Dependent Variable: Ln(Defect) Predictors: Q1 Q2 Q3 Q4 Q5 Q6 Q7 Q8 Q9 Q10 Q11 Q12 Q13 Q14 Q15 Q16 Q17 Q18 Q19 Q20 Q21 Q22 Ln(FP) Ln(Developer) Ln(Duration) | | | |

TABLE 2. Model summary of first-round stepwise linear regression

| Model | R | R Square | Adjusted R Square |
|---|---|---|---|
| 9 | 0.739[i] | 0.546 | 0.523 |
| i. Predictors: (Constant), Ln(FP), Q10, Q11, Ln(Duration), Q3, Q18, Q9, Q2, Q17 | | | |

TABLE 3. Coefficients of first-round stepwise linear regression

| Model | | Unstandardized Coeff. | Standardized Coeff. | Sig. |
|---|---|---|---|---|
| 9 | (Constant) | -2.675 | | 0.001 |
| | Ln(FP) | 0.457 | 0.521 | 0.000 |
| | Q10 | 0.318 | 0.235 | 0.000 |
| | Q11 | 0.216 | 0.159 | 0.008 |
| | Ln(Duration) | 0.449 | 0.216 | 0.000 |
| | Q3 | -0.178 | -0.132 | 0.016 |
| | Q18 | 0.150 | 0.111 | 0.031 |
| | Q9 | 0.183 | 0.135 | 0.023 |
| | Q2 | 0.183 | 0.135 | 0.010 |
| | Q17 | 0.141 | 0.104 | 0.042 |

TABLE 4. Model summary of second-round CATREG

| | Multiple R | R Square | Adjusted R Square |
|---|---|---|---|
| Standardized Data | 0.740 | 0.548 | 0.501 |
| Dependent Variable: Ln(Defect) Predictors: Q2 Q3 Q9 Q10 Q11 Q17 Q18 Ln(FP) Ln(Duration) | | | |

TABLE 5. Coefficients of second-round CATREG

| | Standardized Coefficients | Sig. |
|---|---|---|
| Q2 | 0.140 | **0.057** |
| Q3 | -0.135 | 0.012 |
| Q9 | 0.141 | 0.002 |
| Q10 | 0.225 | 0.000 |
| Q11 | 0.160 | 0.006 |
| Q17 | 0.107 | **0.135** |
| Q18 | 0.113 | 0.024 |
| Ln(FP) | 0.525 | 0.000 |
| Ln(Duration) | 0.214 | 0.000 |

TABLE 6. Model summary of second-round stepwise linear regression

| Model | R | R Square | Adjusted R Square |
|---|---|---|---|
| 9 | 0.740[i] | 0.548 | 0.526 |
| i. Predictors: (Constant), Ln(FP), Q10, Q11, Ln(Duration), Q3, Q18, Q9, Q2, Q17 | | | |

TABLE 7. Coefficients of second-round stepwise linear regression

| Model | | Unstandardized Coeff. | Standardized Coeff. | Sig. |
|---|---|---|---|---|
| 9 | (Constant) | -2.676 | | 0.001 |
| | Ln(FP) | 0.460 | 0.524 | 0.000 |
| | Q10 | 0.306 | 0.226 | 0.001 |
| | Q11 | 0.221 | 0.163 | 0.007 |
| | Ln(Duration) | 0.446 | 0.214 | 0.000 |
| | Q3 | -0.179 | -0.132 | 0.015 |
| | Q18 | 0.152 | 0.112 | 0.028 |
| | Q2 | 0.189 | 0.140 | 0.008 |
| | Q9 | 0.188 | 0.139 | 0.022 |
| | Q17 | 0.147 | 0.109 | 0.033 |

We also listed the coefficient results of CATREG in Table 5. The results still showed that two of the derived coefficients were not statistically significant (p-value > 0.05), which had to be further examined by stepwise linear regression.

Next we applied the second-round stepwise linear regression with the CATREG quantifications for the specific independent variables. The summary of the stepwise regression was presented in Table 6. Once again only the last one (No. 9) of the stepwise steps was listed and the results of the previous stepwise steps were omitted for simplicity. The results confirmed the final model included exactly the 9 predictors, with adjusted R-square being 0.526 and p-value 0. The final R-square was the highest among all the experiments.

The coefficients of the final model were listed in Table 7 and all of them were significant (p-value < 0.05).

Therefore we obtained the final regression formula as follows:

$Ln(Defects) = 0.460*Ln(FP) + 0.446*Ln(Duration) + 0.189*Q2 - 0.179*Q3 + 0.188*Q9 + 0.306*Q10 + 0.221*Q11 + 0.147*Q17 + 0.152*Q18 - 2.676$ (1)

For all the questions, the choices were arranged in a way from weak to strong. When we looked at the coefficients in the formula, only the one of Q3 was negative, which meant that more experienced developers inclined to produce fewer defects. For other questions, i.e. Q2 (release frequency), Q9 (data complexity), Q10 (computational complexity), Q11 (structural complexity), Q17 (bug tracking tool) and Q18 (users involved), the defect trend conformed to the order of the answers, i.e. higher level of the answers would result in more defects. For the two quantitative predictors, both had positive coefficients. Therefore bigger size resulted in more defects and longer duration of development led to more defects. All the findings were consistent with our intuition.

We also verified the predictors in the final formula by other techniques such as correlation analysis and ANOVA.

Moreover, we tried to begin the regression process with various combinations of predictors. The results were consistent, which proved the selection of the independent variables were effective and robust.

We have mentioned that ordinary regression methods are not sound solutions when some predictors are qualitative and with more than two categories. However, a certain recoding approach can be applied and frequently dummy variables are designed to replace the original variable in order to perform traditional regression. Therefore, several dummy variables have to be included in place of each categorical predictor, which makes the final regression formula very complicated. Moreover, the coefficients of the dummy variables are very hard to reason. Lastly, the dummy variables and other predictors are included in the regression, but it is impossible to determine which ones are more significant predictors and should enter the model first.

We still carried out ordinary regression using dummy variables despite its weaknesses discussed above. We wanted to compare the estimation performance of our approach with that of the ordinary regression. In software estimation area, the most widely used evaluation criterion is the mean magnitude of relative error (MMRE) [10]. The magnitude of relative error (MRE) is calculated by:

$$MRE = \frac{|y_i - \hat{y}_i|}{y_i}, \quad (2)$$

where $y_i$ is the actual value and $\hat{y}_i$ is the predicted value.

And therefore, MMRE is computed as:

$$MMRE = \frac{1}{n}\sum_{i=1}^{n} MRE_i. \quad (3)$$

First we processed all data points using ordinary regression. The resulted MMRE was 0.9451. The estimation model derived by our approach obtained a little lower MMRE, 0.9431. The improvement is not significant. A more convincing method for model evaluation is cross-validation, in which the dataset is divided into *k* subsamples. One subsample is reserved as validation (testing) data, while the remaining *k*-1 subsamples are kept as training data for building the model. The cross-validation process is therefore to be repeated *k* times, and the *k* results are averaged to find out the final performance of that model. We conducted a 6-fold cross-validation with MMRE as the evaluation criterion. The results are presented in Table 8. The average MMRE of our approach was 1.2727, compared with 1.4083 of the ordinary regression method. The average improvement was also not obvious, but one experiment (Experiment 4) resulted in a significant improvement. The reason might be that there was unbalance in the data and bias in the division. Similar results were derived when 10-fold cross-validation was conducted. In conclusion, we could claim that our approach would develop a model better than the ordinary regression method using MMRE, without emphasizing on the other merits of our approach.

## V. CONCLUSIONS AND FUTURE WORK

In this paper we performed statistical techniques to build a software quality estimation model based on data from OSS projects.

Two regression methods are suggested to accomplish the task. CATREG is used to acquire optimal quantifications of the qualitative predictors in the meantime the regression process is carried out. With the quantifications from CATREG, stepwise linear regression can be applied to further validate the significance of the predictors. The two steps can be repeated until the final regression formula is derived.

TABLE 8. MMRE results of 6-fold experiments

| MMRE | Regression (Dummy variables) | CATREG + Stepwise | Improvement |
|---|---|---|---|
| Experiment 1 | 1.3690 | 1.3657 | 0.0033 |
| Experiment 2 | 1.6432 | 1.6425 | 0.0007 |
| Experiment 3 | 0.7784 | 0.7747 | 0.0037 |
| Experiment 4 | 1.9725 | 1.1736 | 0.7989 |
| Experiment 5 | 1.5635 | 1.5581 | 0.0054 |
| Experiment 6 | 1.1229 | 1.1216 | 0.0013 |
| Average | 1.4083 | 1.2727 | 0.1356 |

It has been a problematic issue to assign appropriate numeric values (quantifications) to those categorical (qualitative) variables in building software estimation models. In most cases it is done intuitively with subjective judgments. CATREG provides a method to achieve the goal directly from data. Moreover, the predictor selection is robust by using CATREG together with stepwise regression, and the predictors are determined to enter into or exclude from the model during the process. Last but not least, the derived estimation model is easy to interpret and more manageable in software practices.

The suggested techniques were applied to data of OSS projects. The experiments have proved the process is effective and valuable. The final estimation model is not accurate enough to fit the data points perfectly. It is mostly due to the problems of data quality raised by the data collection and the casual characteristics of OSS projects.

Our future work includes the following two aspects:

(1) The quantifications and parameters of the derived model could be further calibrated to achieve better performance. We plan to apply soft computing techniques to accomplish the optimization tasks;

(2) As the only realistic data source consists of projects from the open source community, we plan to develop a way to acquire more quality-related information. When more predictors and more accurate defect information are available, the proposed approach can be used for building a real tool for OSS quality estimation.

ACKNOWLEDGMENT

The authors would like to thank the OSS project administrators who responded to the questionnaire to help us accomplish the research.

## AUTHORS PROFILE

Jie Xu is currently a Ph.D. student with the Electrical and Computer Engineering Department, the University of Western Ontario. His research interests include software engineering, software estimation, soft computing, and data mining. His email address is jxu89@uwo.ca.

Danny Ho is an independent management consultant and advisor to two startup companies. He is also appointed as an Adjunct Research Professor at the Department of Software Engineering, Faculty of Engineering, The University of Western Ontario and University of Ontario Institute of Technology. His areas of special interest include software estimation, project management, object-oriented software development, and complexity analysis. He is currently a member of the Professional Engineers Ontario (PEO) and a Project Management Professional (PMP). His email address is danny@nfa-estimation.com.

Luiz Fernando Capretz is currently an Associate Professor and the Director of the Software Engineering Program at the University of Western Ontario, Canada. His present research interests include software engineering (SE), human factors in SE, software estimation, software product lines, and software engineering education. He is an IEEE senior member, ACM distinguished member, MBTI certified practitioner, Professional Engineer in Ontario (Canada). He can be reached at lcapretz@eng.uwo.ca.

## APPENDIX A. QUESTIONNAIRE

1. Is there a specific plan/schedule for the project?
   ( )A. No schedule
   ( )B. Somehow clear schedule
   ( )C. Very clear schedule

2. How often will the project publish new releases (on average)?
   ( )A. Not sure
   ( )B. Every year
   ( )C. Every six months
   ( )D. Every quarter
   ( )E. Every month
   ( )F. Every week

3. What is the average related software development experience of the developers? (language, application and platform)
   ( )A. <1 year
   ( )B. 1-3 years
   ( )C. 3-5 years
   ( )D. >5 years

4. What is the percentage of personnel change during the development?
   ( )A. <10%
   ( )B. 10% - 20%
   ( )C. 20% - 30%
   ( )D. 30% - 40%
   ( )E. >40%

5. Is there any similar project (functionality and implementation)?
   ( )A. None
   ( )B. A few
   ( )C. Many

6. Is there any reliability requirement for the project?
   ( )A. Low: Slight inconvenience or very small losses when fails;
   ( )B. Nominal: Moderate losses when fails;
   ( )C. High: High financial losses, or risk to life when fails.

7. Is there any response time constraint?
   ( )A. Low: No or loose time constraint;
   ( )B. Nominal: Common time constraints, no special request;
   ( )C. High: Strict time limit or real time system.

8. How do you deal with modularity in the project?
   ( )A. No consideration of modularity
   ( )B. Redesigned during the development stage
   ( )C. Prepared at the beginning of the development phase
   ( )D. Clearly designed during the design stage

9. What is the complexity of data management in the project?
   ( )A. Very low: Simple arrays; Simple DB queries, updates;
   ( )B. Low: Single file with no data structure changes, no edits, no intermediate files. Moderately complex DB queries, updates;
   ( )C. Nominal: Multi-file input and single file output. Simple structural updates. Complex DB queries, updates;
   ( )D. High: Simple triggers. Complex structural updates;
   ( )E. Very high: Distributed or complicated database management. Complex triggers. Search optimization.

10. What is the computational requirement in the project?
    ( )A. Very low: Only basic math expressions involved;
    ( )B. Low: Standard math/statistical routines needed;
    ( )C. Nominal: Basic numerical data analysis like ordinary differential equations and regular calculation accuracy required;
    ( )D. High: Complex data analysis such as partial differential equations;
    ( )E. Very high: Accurate numerical analysis with noisy, stochastic data.

11. What is the level of control flow in the project?
    ( )A. Very low: Straightforward nesting structured programming with simple decision conditions;
    ( )B. Low: Basic nesting with decision tables; Simple callback and message exchange;
    ( )C. Nominal: Highly structured programming with complicated predicates; Queue and stack control; Basic distributed processing;
    ( )D. High: Recursive coding; Simple interrupt handling; Task synchronization, complex callbacks, complex distributed processing; Soft real time control;
    ( )E. Very high: Complex interrupt handling with changing priorities; Immediate real time control.

12. What is the requirement of user interface management?
    ( )A. Low: Simple forms;
    ( )B. Nominal: Graphic user interface;
    ( )C. High: 2D/3D, dynamic graphics; multimedia.

13. Do you have test plan for the project?
    ( )A. No test plan
    ( )B. Somehow clear plan (basic requirements)
    ( )C. Very clear test plan (test phases, test cases)

14. Do you use any tool for testing?
    ( )A. No
    ( )B.Yes (Name ________)

15. What percentage of source code is covered during testing?
    ( )A. < 20%
    ( )B. 20% - 40%
    ( )C. 40% - 60%
    ( )D. 60% - 80%
    ( )E. > 80%

16. The previous coverage information is derived from:
    ( )A. Rough estimation
    ( )B. Coverage tool (Name ________)

17. Is the total number of bugs recorded correctly in the Bug Tracking System?(If not, please give a number)
    ( )A. No (Number ________)
    ( )B.Yes

18. How many users are involved in the project?
    ( )A. < 5
    ( )B. 5 - 10
    ( )C. 10 - 50
    ( )D. 50 - 100
    ( )E. > 100

19. What percentage of defects/bugs do users report?
    ( )A. < 20%
    ( )B. 20% - 40%
    ( )C. 40% - 60%
    ( )D. 60% - 80%
    ( )E. > 80%

20. What percentage of total development effort is used for testing?
    ( )A. < 20%
    ( )B. 20% - 40%
    ( )C. 40% - 60%
    ( )D. 60% - 80%
    ( )E. > 80%

21. What documentation is used to help new developers get onboard?
    ( )A. No particular documentation
    ( )B. Major guidelines available
    ( )C. Detailed definition of processes and development guidelines available

22. How is the user documentation prepared?
    ( )A. No particular documentation
    ( )B. Only draft and incomplete version
    ( )C. Important parts covered
    ( )D. Detailed and comprehensive

End of the questionnaire

APPENDIX B. NORMAL PROBABILITY PLOTS

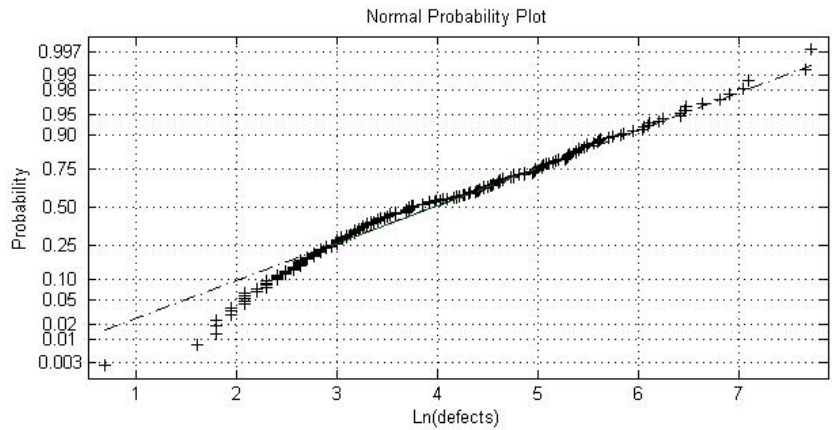

Jie Xu et. al. / (IJCSE) International Journal on Computer Science and Engineering
Vol. 02, No. 06, 2010, 1952-1958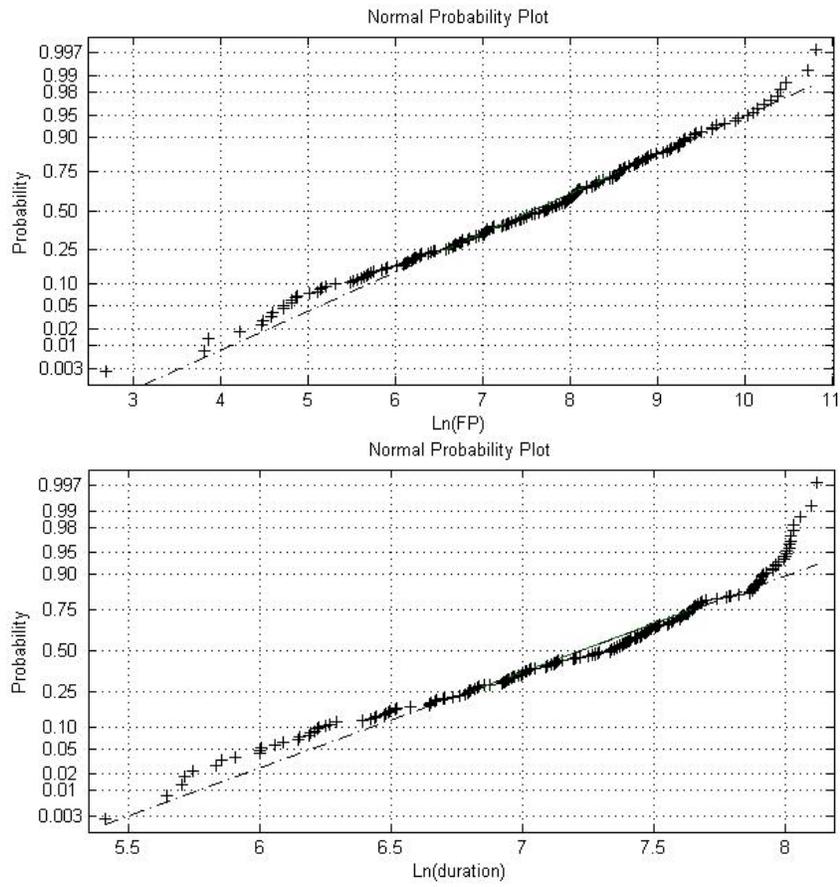

ISSN : 0975-3397　　　　　　　　　　　　　　　　　　　　　　　　　　　　　　　　　　　　　　　　　1958